\documentclass[onecolumn,floatfix,aps,superscriptaddress]{revtex4}
\usepackage{graphicx}
\usepackage{amsmath}
\usepackage[colorlinks=true, citecolor=blue, urlcolor=blue, linkcolor = blue]{hyperref}
\usepackage{amssymb}
\usepackage{bm}
\usepackage{color}
\usepackage{bookmark}
\usepackage{tabularx}
\usepackage{mathtools}
\usepackage{microtype}
\usepackage{cancel}
\usepackage{relsize}
\usepackage{float}
\usepackage{xcolor}

\graphicspath{{../Images/}}

\begin{document}

\title{Novel curved   solitons of integrable (2 +1) dimensional KMN equation}

\author{Abhik Mukherjee}
\affiliation{Theoretical Physics and Quantum Technologies Department, NUST ``MISIS'', Moscow, Russia}


\begin{abstract}
In this letter, the unique exact lump and topological soliton solutions of integrable (2+1) dimensional Kundu-Mukherjee-Naskar (KMN) equation are obtained.  These solutions have an unusual property that they can get curved in the $x-y$ plane arbitrarily due to the presence of an arbitrary function of space($x$) and time($t$) in their analytic forms. Due to this  special feature, the solutions can be used to model the bending of optical solitonic beam, different types of wave structures in real physical experimental conditions. This novel feature, which is a rare property for a constant coefficient completely integrable equation, arises due to the Galilean co-variance property and ``current like'' nonlinearity present in the KMN equation.
\end{abstract}

\maketitle

\section{Introduction}
  Line soliton solutions which are exact stable localized solutions of completely integrable systems,  have been explored extensively in the past few decades \cite{Johnson,Lakshmanan}. However, in recent years, a deep research interest has been developed on some special intricate solutions like accelerating solitons \cite{accsol, Kundu-PLA-2019, Zaqilao},
topologically nontrivial solutions \cite{toposol, TopB1, TopB2}, rogue wave solutions  \cite{ourpaper, Rogue} etc. Recently a  new completely integrable (2+1) dimensional nonlinear evolution equation has been derived by Anjan Kundu, Abhik Mukherjee and Tapan Naskar to describe the dynamics of two-dimensional oceanic rogue wave phenomena \cite{ourpaper}. This  equation was also re-derived in modeling nonlinear ion acoustic waves in magnetized plasma system \cite{Kundu-PoP-2015}. Using triplet Lax pair, the first order rogue wave solution of this equation is obtained in \cite{CNSNS-He-2016} using one fold Darboux Transformation. Also, the authors have named this nonlinear dynamical equation as Kundu-Mukherjee-Naskar (KMN) equation in their paper \cite{CNSNS-He-2016}. Thereafter, a substantial amount of research has been carried out on exploring different kinds of exact solutions of KMN equation like higher order rational solutions \cite{Roman-2017}, various kinds of optical soliton solutions  having different features \cite{Biswas1-2020,CJP-2019,Optik-2019a,Optik-2019b,Rizvi,Sulaiman,Aliyu,Jalili,Khater,Sudhir1}, general solutions \cite{Optik-2019c}, power series solutions  \cite{MPLB-2018}, complex wave solutions \cite{Jhangeer}, periodic solutions (via variational principle) \cite{J-He}, solitons in birefringent fiber system \cite{Optik-2KMN-2019,Optik-2KMN-2019a,Yildirim-f} etc. 
Complete integrability features including the existence of Lax pairs, conservation laws, higher soliton solutions via Hirota method, symmetry analysis, nonlinear self-adjointness property of KMN equation are explored in \cite{ourpaper,Sudhir2}. One of the interesting features of KMN model is that it has both bright as well as dark soliton solutions \cite{Sudhir1} irrespective of it's  dispersive term that is different from the standard NLS equation that admits only bright (dark) soliton for focusing (defocusing) nonlinearity. Such feature comes from the current like nonlinearity (that depends on both wave envelope and it's spatial derivative) present in KMN equation.

In real physical situations or experimental conditions, the localized wave structure describing surface wave perturbations or density fluctuations may have various shapes. In that case, the application of standard soliton solutions having constant amplitude and velocity in modeling such localized wave phenomena may be restricted because the real wave structure may twist, turn or bend due to external disturbances. In this letter, we will discuss about some special  solutions of KMN equation having rare analytic beauty. The equation admits special lump and topological soliton solutions that can get curved in $x-y$ plane arbitrarily due to presence of an arbitrary function of space($x$) and time($t$) in their analytic expressions. Generally, solitary wave solutions with variable velocities arise  in in-homogeneous media \cite{Xiang}, in equations with  variable  coefficients \cite {Liu,OE,AML}.  Thus this novel bending feature is rare for an integrable equation having constant coefficients. This special property of those localized solutions arise from the Galilean co-variance property and current like nonlinearity present in KMN equation which  will be discussed in this brief paper. Arbitrary bending of optical solitonic beam has been discussed before in \cite{KunduNaskar} where the boundary value of population inversion of the medium becomes non uniform. In this work, we will show such arbitrary bending of solitons using completely integrable (2+1) dimensional KMN model having constant coefficients.

 \section{Galilean co-variance (GC)}
 We start our analysis from the novel nonlinear integrable (2 + 1) dimensional  Kundu-Mukherjee-Naskar (KMN) equation   given as 
\begin{equation}
 iu_t + u_{xy} + 2iu \ (u u_{x}^* - u^*u_{x}) = 0, \label{KMN}
\end{equation}
where $u(x,y,t)$ is the wave envelope, asterisk sign denotes complex conjugation and the subscripts denote partial derivatives with respect to space ($x, y$) and time ($t$) coordinates.
The equation (\ref{KMN}) remains co-variant as 
\begin{equation}
 i U_T + U_{XY} + 2iU \ (U U^*_{X} - U^*U_{X}) = 0, \label{KMN2}
\end{equation}
 under the Galilean transformation:
 \begin{eqnarray}
  X = x - a t, \ \
  Y = y, \ \
  T = t, \ \
  u(x, y, t) = U(X, Y, T) \ e^{i a Y}, \label{GT}
 \end{eqnarray}
where $a$ is the constant frame velocity. This co-variance property is important to find the time-dependence  of the curved solitons which will be discussed  in the following sections.

\section{Curved lump solitons}

Rogue waves (RWs) are deep,  nonlinear, oceanic surface waves \cite{ourpaper} having applications  in various physical systems  \cite{Rogue}. 
The most popular model of single lump RW in (1+1) dimensions is the famous Peregrine breather solution \cite{Rogue} of Nonlinear Schrodinger Equation (NLSE). Though it is vastly  used to model various RW phenomena,   yet the maximum amplitude of the solution remains fixed
at a certain value (3 times of background wave) due to  the absence of  any free parameter in the solution \cite{Rogue}.
The single peak RW solution of other equations like Kundu Ekkaus equation  \cite{kunduekkaus}, Davey Stewartson system \cite{DS}, Hirota equation \cite{Hirota} etc are also devoid of free parameters.  The free parameters  arise in higher order multiple peak rogue wave solutions having more complicated analytic structures \cite{HPB1,HPB2,HPB3}.
Hence there exists a necessity  in finding a possible  RW solution in (2+1) dimensions which would allow  free parameters and arbitrary functions in their expressions to regulate amplitude, width and shape of the wave phenomena. The single peak lump soliton solution  of KMN equation (\ref{KMN}) is derived in \cite{ourpaper} having three free parameters to control RW dynamics by including the effects of ocean current.

We  derive in this letter,  an exact   lump soliton describing RW phenomena  from the KMN equation (\ref{KMN}) as
\begin{equation}
 u(x, y, t) = e^{i(4  + a)y} \ [-1 + \frac{(1 - 4 i y)}{c+ 4 y^2 + \alpha f(X)^2 }],
\label{rog}\end{equation}
where $f(X)$ is an arbitrary function of $x, t$ with $X = (x - a t).$
The modulus of the rogue wave solution (\ref{rog}) 
would go to a nontrivial value at space  infinities $(x, y \rightarrow \pm \infty)$ like the Peregrine breather solution for suitable choice of $f(X)$. It  is important to note that this RW solution (\ref{rog}) includes three  free
parameters $a, \alpha$, $c$ and an arbitrary function $f(X)$, with $X = (x - a t)$ that would be important in modeling the real RW  dynamics. Also, due to the presence of the arbitrary  function of $x, t$ : $f(X)$,  the lump solution (\ref{rog}) can get curved in the $x-y$ plane arbitrarily which is a unique and novel feature for  a completely integrable equation with constant coefficients. The nature of curvature  would depend on the choice of function $f(X)$ as shown in FIG. 1.
Thus we have the liberty to generate different wave structures from a single solution (\ref{rog}) by choosing different functional forms of $f(X); X = x - a t$.
The condition which is crucial for the sudden appearance of RWs modeled by (\ref{KMN})  is the modulation instability that has been discussed in \cite{ourpaper}. Whatever be the analytic forms of the arbitrary function $f(X); X = (x - a t)$, it gets cancelled in the equation (\ref{KMN})   due to $GC$ (\ref{GT}) and the current like nonlinearity present in (\ref{KMN}).

\begin{figure}
\includegraphics[width=4cm]{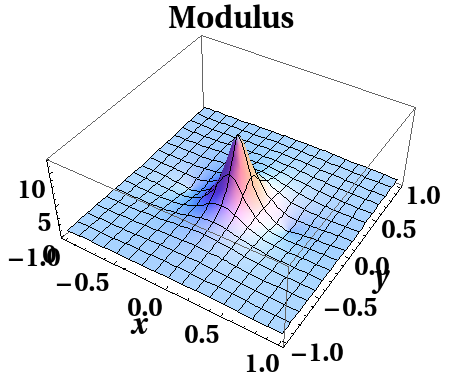}
 \ \ \ \includegraphics[width=4cm]{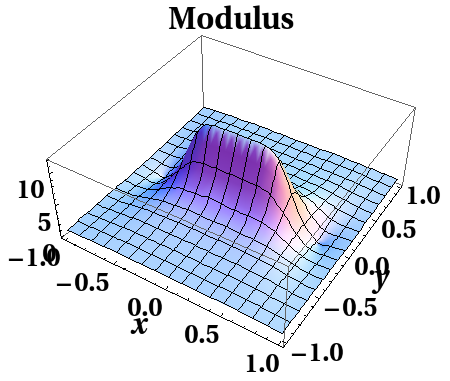}
\ \ \ \ \ \ \includegraphics[width=4cm]{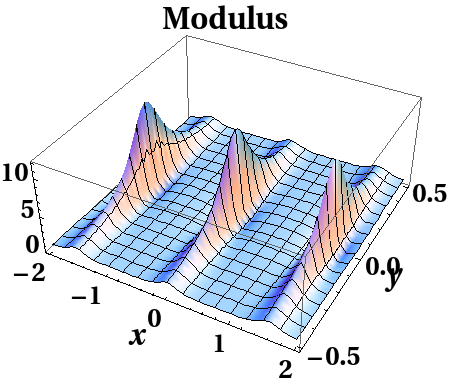}

 (a) $f(X) = X = (x - a t)$
\quad \qquad \qquad  (b) $
f(X) = X^2 = (x - a t)^3$
\quad \qquad \qquad  (c) $f(X) = \sin{2X} = \sin{2(x - a t)}$
 
\vspace{0.5cm}

\noindent FIG.1: { \bf We can see from above figures (FIG. 1(a) - FIG. 1(c)) that the RW solution (\ref{rog})   gets curved (distorted) in the $x-y$ plane for the specified choice of the arbitrary function $f(X)$. This is a unique feature for KMN equation (\ref{KMN}) which allows us to chose $f(X)$ arbitrarily to generate different wave structures from the same solution (\ref{rog}).  Constants in the plots are chosen as $\alpha_1 = 4, c = \frac{1}{13}, a = 1, P = 1$.    }
\end{figure}

\section{Curved topological solitons}
There has been a continued
interest  in the field of fascinating  topological soliton solutions \cite{toposol,TopB1,TopB2} both in physics and mathematics. 
Exact one topological line soliton solution with constant amplitude and velocity of the KMN equation (\ref{KMN}) is derived in \cite{Sudhir1}.
In this letter, we will
explore the possibility of finding the exact curved  topological soliton solutions
of  (\ref{KMN}) which would be rather an interesting problem
due to the homogeneity of the equation with constant coefficients.  We start from the form (\ref{KMN2}) of the equation by taking the following ansatz :
\begin{align}
&{} U (X, Y, T) = R (\xi) \  e^{i \theta}, \nonumber \\
&{} \xi = \alpha_1 X + \beta_1 Y + \gamma_1 T + A(X), \ \  \theta = \alpha_2 X + \beta_2 Y + \gamma_2 T + B(X), \label{ansatz} 
\end{align}
where $\alpha_i, \beta_i, \gamma_i, (i= 1,2)$ are constant parameters and $A(X), B(X)$ are functions of X. Applying the ansatz (\ref{ansatz}) on Eq. (\ref{KMN2}) and collecting real and imaginary parts we get the following two equations :
\begin{align}
&{}-R \theta_T + R_{XY} - R \theta_X \theta_Y + 4 R^3 \theta_X = 0, \label{Re} \\
&{}R_T + R_X \theta_Y + R_Y \theta_X + R \theta_{XY} = 0, \label{Im}
\end{align}
where the subscripts denote partial derivatives. From (\ref{Im}) we get 
\begin{equation}
 R_{\xi} \gamma_1 + R_{\xi} (\alpha_1 + A_X) \beta_2 + R_{\xi} (\alpha_2 + B_X) \beta_1 = 0. \label{ABrel}
\end{equation}
If we choose $\gamma_1 = 0,$ then we can get a relation between the two arbitrary functions $A(X)$  and $ B(X)$ as 
\begin{equation}
 \beta_2 (\alpha_1 + A_X) = - \beta_1 (\alpha_2 + B_X). \label{ABrel2}
\end{equation}
Similarly from (\ref{Re}) we can obtain 
\begin{equation}
 -R \gamma_2 + R_{\xi \xi} \beta_1 (\alpha_1 + A_X) - R \beta_2 (\alpha_2 + B_x) + 4 R^3 (\alpha_2 + B_X) = 0. \label{gam20}
\end{equation}
It is important to note that, if we choose $\gamma_2 = 0$ and use the relation (\ref{ABrel2}) in (\ref{gam20}) we would get a constant coefficient ordinary differential equation (ode) given as
\begin{equation}
 R_{\xi \xi} + R -\frac{4}{\beta} R^3 = 0, \label{ode2}
\end{equation}
where we have chosen $\beta_1 = \beta_1 = \beta$ for mathematical convenience which does not violate the
 generality of the problem. Hence, by choosing $\gamma_1 = \gamma_2 = 0$, we get a constant coefficient ode (\ref{ode2})  from (\ref{Im}) in spite of the presence of arbitrary functions $A(X), B(X)$ in the solution (\ref{ansatz}). This is a rare feature which is present in the KMN equation (\ref{KMN2}) due to it's current like nonlinearity (nonlinear term is a function of $U$ and $U_X$). Then, for a special condition ($\gamma_1 = \gamma_2 = 0$) the arbitrary functions present in (\ref{gam20}) get cancelled from both sides of equation. In case of
 standard NLSE, such cancellation is not possible due to it's amplitude like nonlinearity (nonlinear term is a function of $U$ only, not its derivatives). So this is a unique and novel feature present in our equation (\ref{KMN2}) in spite of it's complete integrability. We will find one soliton solution for the sake of simplicity though it can be solved to express via Jacobi elliptic function. Now solving (\ref{ode2}) we get 
 \begin{equation}
 R =  P \tanh{[\nu \{\alpha_1 X + A(X) + \beta Y \}]},  \ \nu = \pm \frac{1}{\sqrt{2}}, \ \beta = 4 P^2.  \label{R}
 \end{equation}
Hence, we can get  the exact curved (due to presence of an arbitrary function $A(X)$) one topological soliton as 
 \begin{align}
&{}U(X, Y, T) = \pm P \tanh{[\frac{1}{\sqrt{2}} \{\alpha_1 X + A(X) + \beta Y \}]} \  e^{i [-\alpha_1 X - A(X) + \beta Y + c_0]}, \ \beta = 4 P^2 , \label{top1}
 \end{align}
where $P, c_0$ are free parameters. Back boosting to the original frame $(x - y - t)$ we get the final expression of the curved soliton from (\ref{top1}) as
\begin{align}
&{}u(x, y, t) = \pm P \tanh{[\frac{1}{\sqrt{2}} \{\alpha_1 X + A(X) + 4 P^2 y \}]} \  e^{i [-\alpha_1 X - A(X) + (4 P^2 + a) y + c_0]}, \label{top2}
 \end{align}
where $X = x - a t$ with $a$ being the constant frame velocity. Thus in spite of choosing $\gamma_1 = \gamma_2 = 0$, we get the time dependence in (\ref{top2}) (since, $X = x - a t$) due to the GC property (\ref{GT}). Now choosing different functional forms of $A(X); \ X = x - a t,$ we can get different curvatures of the one topological soliton solutions (\ref{top2}) in the $x - y$ plane which are shown in FIG. 2.
The appearance of the
arbitrary function $A(X) ; X = x - a t,$ in the exact dark soliton solution (\ref{top2}) is a unique feature of our equation (\ref{KMN}). This feature  differs drastically from other  nonlinear evolution equations of NLSE and
derivative NLSE family.

\begin{figure}

\includegraphics[width=4cm]{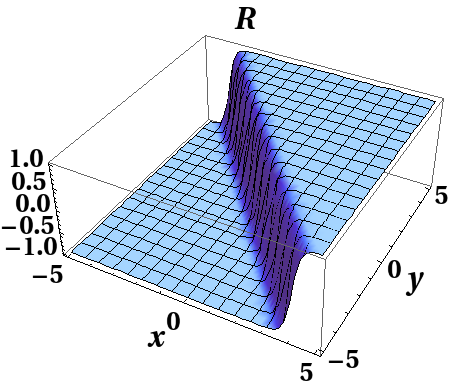}
 \ \ \ \includegraphics[width=4cm]{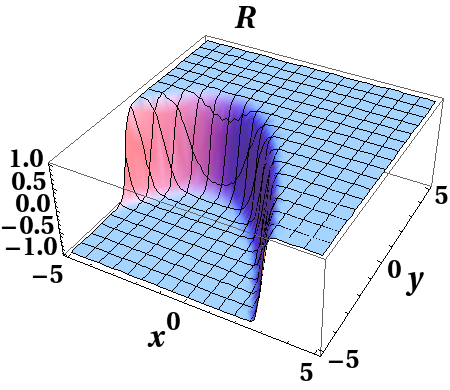}
\ \ \ \ \ \ \includegraphics[width=4cm]{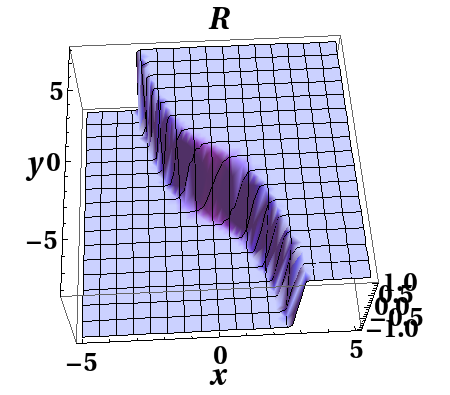}

 (a) $A(X) = X = (x - a t)$
\quad \qquad \qquad  (b) $
A(X) = X^2 = (x - a t)^2$
\quad \qquad \qquad  (c) $A(X) = X^3 = (x - a t)^3$
 
 \vspace{0.5cm}

\noindent FIG.2: {\bf We can see from above figures (FIG 2(a) - FIG. 2(c)) that the one topological soliton solution (\ref{top2})  gets curved in the $x-y$ plane for the specified choice of the arbitrary function $A(X)$. For the sake of simplicity, we have plotted the function $R (x, y, t)$ (\ref{R}) in the $x-y$ plane. This is a unique feature for KMN equation (\ref{KMN}) which is an integrable equation with constant coefficients.  Constants in the plots are chosen as $ a = 1, P = 1, c_0 = 1.$    }
\end{figure}

We can also see  that the
freedom of choice of the arbitrary function $A(X) ; X = x - a t,$  can also lead to solitons with changing topological properties.
We can explain this feature by introducing a new quantity $C$  which is defined through the boundary condition
as
\begin{equation}
 C = \frac{1}{2P} [ R(x\to \infty) - R(x\to -\infty)]. \label{topcharge}
\end{equation}

\begin{enumerate}
 
\item 
If we choose  $A(X)$ = $X$ = $(x - a t)$, then 
 we get a typical  kink or antikink line soliton  which is linked to  $C = \pm 1$. This can be understood from the expression of $R$ in (\ref{R}). It is interesting to see that, we can obtain multi-kink, antikink solutions etc  from the same form of $R$ in (\ref{R}) for different choice of $A(X)$. This will change the
 topological properties of the soliton solution
which is a  rare phenomenon in soliton theory. 

\item 
For $A(X)= X^2 = (x - a t)^2$, the
topological characteristic becomes trivial since 
$C = 0 $ due to the square term in the argument of the solution (\ref{R}). This trivial value of $C$ can be explained by the simultaneous appearance of kink and antikink due to this particular choice of $A(X)$.
In case of standard NLSE, the expressions of kink-antikink soliton solutions are more complicated.

\item 
For $A(X)= X^3 = (x - a t)^3$ we would get again
$C = \pm 1 $, which can also be understood from the expression of $R$ in (\ref{R}). This choice shows the  generation of kink, antikink and kink solitons from the same solution (\ref{R}).

\end{enumerate}

In the same way we can generate different types of curvature in waves from the solutions (\ref{rog}) and (\ref{top2}) by choosing different functional forms of the arbitrary functions present in their analytic expressions. This would be useful in modeling the bending phenomena of optical solitonic beam \cite{KunduNaskar}.
The bending of an optical beam was obtained in
 \cite{B1,B2} through an Airy
function solution of linear Schrodinger equation that could preserve
its parabolic form over a finite distance. The beam acceleration along an arbitrary curve was achieved in \cite{B3} though at the cost of
non-preservation of the shape. In contrast to these research work, our bending features are more general preserving it's constant amplitude which may be interesting in both theoretical and experimental physics.

\section{Conclusions}
We can conclude our brief paper by stating the fact that we have been able to find the special exact lump and topological soliton solutions of integrable (2+1) dimensional KMN equation (\ref{KMN}). These solutions can get curved in the $x-y$ plane due to the presence of an arbitrary function of $x,t$ in their analytic forms. This is a rare feature for a constant coefficient completely integrable evolution equation. The bending property originates from the current like nonlinearity present in the equation (\ref{KMN}) and the Galilean co-variance (\ref{GT}). Further research on these novel localized solutions can be carried out exploring it's deep mathematical structures. This may pave new direction of research in nonlinear optics and fluid dynamics. 

\section {ACKNOWLEDGMENTS}

The work has been carried out with partial financial support from the
Ministry of Science and Higher Education of the Russian Federation in the
framework of Increase Competitiveness Program of NUST$ «MISiS» ( K4-2018-061),$ implemented by a governmental decree dated 16 th of March 2013, N 211. Author is grateful to  late Prof. Anjan Kundu for introducing such interesting concept in in this research field. Author also acknowledges Shakya Mukherjee and Ahoban Mukherjee for the fruitful discussions during the progress of the work.


\section{References}

 \begin{thebibliography}{99}
\bibitem{Johnson}
R.S.Johnson, A Modern Introduction to the Mathematical Theory of Water Waves,
Cambridge University Press,Cambridge, 1997.
\bibitem{Lakshmanan} 
M.Lakshmanan, S. Rajasekar, Nonlinear Dynamics: Integrability, Chaos and Patterns, Springer, Berlin,
2003.
\bibitem{accsol}
A.V.Gorbach, D.V Skryabin, Phys.Rev.A 76 (2007) 053803. 
\bibitem{Kundu-PLA-2019}
A. Mukherjee, A. Kundu,  Phys Lett A 383 (2019) 985
\bibitem{Zaqilao}
Zaqilao,  Nonlinear Dynamics 99 (2020) 2945
\bibitem{toposol}
M.O.Katanaev, Theoretical and Mathematical Physics 132(2) (2004) 163.
\bibitem{TopB1}
A. Biswas, D. Milovic, Int. J Theor. Phys 48 (2009) 1104.
\bibitem{TopB2}
A. Biswas, Commun Nonlinear Sci Numer Simulat 14 (2009) 2845.
\bibitem{ourpaper}
A.Kundu, A.Mukherjee, T.Naskar, Proc. Soc. A 470 (2014) 20130576.
\bibitem{Rogue}
C.Kharif, E.Pelinovsky, A.Slunyaev, Rogue Waves in Ocean, Berlin, Germany, Springer, 2009.
\bibitem{Kundu-PoP-2015}
A. Mukherjee, M. S. Janaki, and A. Kundu,  Phys Plasmas 22 (2015) 072302.
\bibitem{CNSNS-He-2016}
D. Qiu,Y. Zhang, J. He,  Commun Nonlinear Sci Numer Simulat 30 (2016) 307-315.
\bibitem{Roman-2017}
X. Wen,  Proc Romanian Acad A, 18 (2017) 191-198.
\bibitem{Biswas1-2020}
A. Biswas et.al, Results in Physics 16 (2020) 102850.
\bibitem{CJP-2019}
M. Ekici, A. Sonmezoglu, A. Biswas, M. R. Belice, Chin J Phys  57 (2019) 72-77.
\bibitem{Optik-2019c}
N. A. Kudryashov, Optik 186 (2019) 22-27. 
\bibitem{Jhangeer}
A. Jhangeer et.al, Results in Physics 16 (2020) 102816.
\bibitem{J-He}
J. H. He , Results in Physics 17 (2020) 103031.
\bibitem{Optik-2019a}
Y. Yildirim,  Optik, 183 (2019) 1061.
\bibitem{Optik-2019b}
Y. Yildirim,  Optik, 184 (2019) 247.
\bibitem{Optik-2KMN-2019}
Y. Yildirim,  Optik 184 (2019) 121.
\bibitem{Optik-2KMN-2019a}
Y. Yildirim,  Optik, 183 (2019) 1026.
\bibitem{Yildirim-f}
Y. Yildirim, M. Mirzazaden, Chin J Phys  64 (2020) 183.
\bibitem{MPLB-2018}
W-Q. Peng, S-F. Tian, Tian-Tian Zhang,  Mod Phys Lett B 32 (2018) 1850336.
\bibitem{Rizvi}
W-Q. Peng, S-F. Tian, Tian-Tian Zhang,  Mod Phys Lett B 34 (2020) 2050074-1.
\bibitem{Sulaiman}
T.A Sulaiman, H. Bulut, Applied Mathematics and Nonlinear sciences 4(2)(2019)513
\bibitem{Aliyu}
A.I. Aliyu, Y. Li, D. Baleanu , Chin J Phys  63 (2020) 410.
\bibitem{Jalili}
R.A Talarposhti et.al , Int. Journal of Modern Physics B, 34 (2020) 2050102
\bibitem{Khater}
H.Gunerhan,  Mod Phys Lett B  (2020) 2050225-1.
\bibitem{Sudhir1}
S.Singh, A. Mukherjee, K.Sakkaravarthi, K.Murugesan (2020) arxiv: 2001.06766v1
\bibitem{Sudhir2}
S.Singh, R. Sakthivel, M.Inc, A. Yusuf , K.Murugesan, Mod Phys Lett B 34 (2020) 2050068.
\bibitem{KunduNaskar}
A. Kundu, T. Naskar, Physica D 276 (2014) 21
\bibitem{kunduekkaus}
X.Wang, B.Yang, Y.Chen, Y. Yang, Phys.Scr.89 (2014) 095210.
\bibitem{DS}
Y. Ohta, J.Yang, Phys.Rev.E 86 (2014) 036604.
\bibitem{Hirota}
A.Ankiewicz, J.M.Soto-Crespo, N.Akhmediev, Phys.Rev.E 81 (2010) 046602.
\bibitem{HPB1}
N.Akhmediev, A.Ankiewicz, J.M Soto Crespo, Phys. Rev  E 80 (2009) 026601. 
\bibitem{HPB2}
P. Dubard,  V.B Matveev, Nat Hazards Earth Syst.Sci, 11 (2011) 667. 
\bibitem{HPB3}
 A. Ankiewicz, D.J Kedziora,  N. Akhmediev, Phys. Lett A 375 (2011)  2782. 
\bibitem{OE}
X.Y.Gao, Ocean Engineering 96 (2015) 245. 
\bibitem{AML}
X.Y.Gao, Applied Mathematics Letters 73 (2017) 143 . 
\bibitem{Xiang}
 C. Xiang,  Applied Mathematics and Computation 216 (2010) 2235. 
 \bibitem{Liu}
 Y.Liu,  Applied Mathematics and Computation 217 (2011) 5866. 
 \bibitem{B1} G.A. Siviloglou, D.N. Christodoulides, 
Opt. Lett. 32 (2007) 979.
\bibitem{B2}
G.A. Siviloglou, J. Broky, A. Dogariu, D.N. Chistodoulides, Phys. Rev. Lett. 99 (2007) 213901.
\bibitem{B3}
E. Greenfield, M. Segev, W. Walasik, O. Raz,  Phys. Rev. Lett. 106 (2011) 213902.
\end{thebibliography}
\end{document}